\documentclass[twocolumn,amsmath,amssymb]{revtex4}
\usepackage{graphicx}
\usepackage{dcolumn}
\usepackage{amsmath}
\usepackage{bm}
\DeclareGraphicsExtensions{.pdf,.png,.jpg}
\usepackage{epstopdf}

\begin{document}

\title{Electronic structure, lattice dynamics and magnetism of new ThXAsN (X=Fe,Co,Ni) superconductors: A First Principles Study}
\author{Smritijit Sen$^1$ and Guang-Yu Guo$^{1,2}$}
\affiliation{Department of Physics and Center for Theoretical Physics, National Taiwan University, Taipei 10617, Taiwan.$^1$
\\Physics Division, National Center for Theoretical Sciences, Hsinchu 30013, Taiwan.$^2$}

\date{\today}
\begin{abstract}
In this work, we present a comparative first principles study of mechanical properties,
electronic structure, phonon dispersion relation, electron-phonon coupling and magnetism
in three isostructural superconductors, namely, ThFeAsN, ThCoAsN and ThNiAsN.
Experimentally, ThFeAsN and ThNiAsN show superconducting properties, while ThCoAsN has not been synthesized.
Our calculated elastic constants show that all these systems are mechanically stable.
Significant differences in the electronic structures of these three compounds
in terms of density of states, band structures and Fermi surfaces, are found.
Our phonon calculations reveal that all the systems including ThCoAsN, are dynamically stable.
Phonon dispersion relations indicate that the optical modes of all the three systems are almost the same
while there are significant variations in the low frequency manifold consisting of mixed modes.
The electron-phonon coupling constants and superconducting transition temperatures
calculated based on the Eliashberg formalism,
predict a rather high $T_c$ of 6.4 K for ThCoAsN and also a $T_c$ of 3.4 K for ThNiAsN which agrees
well with the experimental value of 4.3 K.
Nevertheless, we find a $T_c$ of 0.05 K for ThFeAsN, which is much smaller than the experimental $T_c$ of $\sim$30 K.
However, a simple analysis considering the amplifying effects of spin density wave order and
out-of-plane soft phonon modes suggests that the $T_c$ could be increased considerably to $\sim$10 K.
Finally, we also discuss the effect of anion As height on the electronic structures
and study possible magnetic states in these three compounds.
\vspace{1pc}
\end{abstract}
\maketitle

\section{INTRODUCTION}
Since the discovery of high temperature superconductivity in LaFeAs(O, F) \cite{Kamihara2008}, a large number of new Fe-based superconductors (SCs),
popularly known as pnictides, have been found. Most of the Fe-based SCs are generally categorized into six families depending upon their
chemical compositions, namely 11, 111, 122, 122$^*$, 1111 and 42622 (some times denoted as 21311) \cite{Stewart2011}.
Interestingly, two newly discovered families 112 and 1144 also show some unique features which are not
of usual characteristics of Fe-based SCs \cite{Katayama2013,Iyo2016}.
All these systems have a tetragonal crystal structure with the FeAs/FeSe layers
as the basic building blocks which seem to play an important role in superconductivity.
Along with these Fe-based SCs, some isostructural Ni/Co-based iso-structural systems with or without superconducting properties
have also been reported \cite{Watanabe2007,Watanabe2008,Ronning2008,Tomioka2009}.
For example, LaNiAsO, a 1111 type nickel-based pnictide, was  reported to become superconducting below temperature of 2.75 K~\cite{Watanabe2008}.
On the other hand, LaCoAsO was found to be a ferromagnet with no trace of superconductivity.~\cite{Sefat2008}
Similarly, BaCo$_2$As$_2$, which is isostructural to Fe-based superconductor BaFe$_2$As$_2$, was  also reported to be
non-superconducting~\cite{Sefat2009}.

Apart from its unconventional superconducting property, Fe-based SCs display
a number of exotic normal state properties such as spin density wave (SDW), orbital density wave (ODW),
structural phase transition, and nematic order, which are distinctly different from other classes
of high superconducting transition temperature ($T_c$) SCs including
cuprates~\cite{Dong2008,Cvetkovic2009,2Cvetkovic2009,Fernandes2014,Fernandes2010}.
Although nickel-based pnictides possess the same crystal structure as that of the Fe-pnictides,
the superconducting pairing mechanism in these systems may be quite different from that of the Fe-based SCs.
In particular, $T_c$ of nickel-based pnictides can be calculated rather accurately within the Eliashberg formalism
of the electron-phonon coupling mediated superconductivity \cite{Subedi2008, 2Subedi2008} [see also Sec. III (D) below].
Recent studies show that the role of electron-phonon coupling in Fe-based SCs has been understated especially
in the presence of the SDW type antiferromagnetic order \cite{Coh2016, Li2012, Deng2009}.

Being multi-orbital nature, Fermi surface topology of Fe-based SCs is quite
different from other high $T_c$ SCs and is sensitive to temperature, pressure and impurity doping.
Sign changing $s^{\pm}$ superconducting order parameter is widely
accepted for describing various physical properties of Fe-based SCs \cite{Mazin2008}
but there are strong  evidences of deviation of $s^{\pm}$ pairing scenario \cite{Kuroki2008}.
Therefore, pairing symmetry of these SCs is still under considerable debate \cite{Chubukov2008,Kazuhiko2008,Hirschfeld2011}.
In general, Fe-based SCs have high critical current density as well as high upper critical
magnetic field ($H_{c2}$). Added to that they possess moderate anisotropy (important for transport of charge)
and ductility which give these SCs edge over high $T_c$ cuprates
as far as superconducting applications are concerned \cite{Pallecchi2015}. These Fe-based SCs have all the potential to
be the next generation SCs. Therefore, a better theoretical understanding of these systems
will accelerate the development of more sophisticated technologies based on pnictides SCs.

Superconductivity in the stoichiometric ThFeAsN compound in ambient pressure with remarkably high $T_c$ of 30 K was recently reported \cite{Wang2016}.
This new 1111-type FeAs-based compound with a ZrCuSiAs-type structure, where the (LaO) layers in LaOFeAs are replaced by the ThN layers.
Unlike ReFeAsO series, this newly discovered Fe-based SC does not show any long range magnetic order.
However, evidence of strong magnetic fluctuations above 35 K was found via muon-spin rotation/relaxation and nuclear magnetic resonance techniques \cite{Shiroka2017}.
ThFeAsN has a distinctively shorter $c$-axis (the $a$-axis is similar) as compared to the LaFeAsO series of Fe-based SCs.
This suggests that ThFeAsN possesses an in-built uni-axial chemical pressure along the $c$-axes which has significant impact
in the physical properties observed in these compounds.
The advent of superconductivity in ThFeAsN, without intrinsic doping nor external pressure, and in the absence of long range magnetic order,
makes this system quite unique as compared to the other 1111 FeAs-based SCs.
Therefore, study of low energy electronic structure of ThFeAsN will provide a wealth of information.

Recent discovery of room temperature superconductivity in a number of compounds
including solidified H$_2$S under hydrostatic pressure and its close proximity to electron phonon couplings,
challenge the general consensus of high temperature superconductivity \cite{Errea2015,Li2014}.
It is also shown that superconductivity in structurally similar compound ThNiAsN
is phonon-mediated and superconducting transition temperature estimated from density functional perturbation theory calculations
within the Eliashberg formalism is 3.5 K, which agrees well with the experimentally measured
$T_c$ of 4.3 K \cite{Wang2017,Yang2018}. Therefore, it is important to study the role of electron-phonon coupling
in superconductivity of ThFeAsN.

A recent first principles investigation on ThFeAsN reveals a stripe antiferromagnetic (AF) ground state,
on contrary to the experimental ground state with no long range magnetic order \cite{2Wang2016,Albedah2017}.
Recent theoretical studies which include the effect of the SDW type AF orders
on electron-phonon coupling for various Fe-based SCs such as FeSe and LiFeAs were able to explain the
experimental $T_c$ of these systems as well as its pressure dependence~\cite{arXiv}.
From these works, it is quite evident that one should consider the combined effect of
the AF order and soft out-of-plane lattice vibration on electron-phonon coupling
to understand the observed $T_c$.
Recent electronic structure calculations also disclose that ThFeAsN is a semimetal and possess
a partially nested electron and hole Fermi surface pockets \cite{Singh2016}.
On the other hand, latest theoretical studies \cite{Yang2018} indicate that ThNiAsN possesses
a quite different electronic structure from that of ThFeAsN with no magnetic ordering.
In this work, we also study the mechanical properties, electronic structure and lattice dynamics
of isostructural compound ThCoAsN. A comparative study of ThXAsN (X=Fe,Co,Ni) will help to elucidate
mechanism of superconductivity in these compounds and this is the main goal of this paper.

The rest of this paper is organized as follows. In the next section, we briefly describe
the crystalline structure of ThXAsN (X=Fe,Co,Ni) and also the details of our first-principles calculations.
In Sec. III.A, we first calculate the elastic constants of ThXAsN (X=Fe,Co,Ni), which show that these
systems are mechanically stable. We also discuss some of the important mechanical properties
of these systems from the application point of view.
In addition, we evaluate the plasma frequencies of the systems along different axes
to understand the effective dimensionality
of these systems and compare these values with others Fe-based SCs in Sec. III.B.
Next we report the calculated electronic structures of ThXAsN (X=Fe,Co,Ni), especially
density of states (DOS), band structure and Fermi surface in Sec. III.C.
Then, in Sec. III.D, we present the calculated phonon dispersion relations and electron-phonon coupling for ThXAsN (X=Fe,Co,Ni).
We also evaluate superconducting transition temperatures of these compounds within the Eliashberg theory.
In Sec. IV, we study possible magnetic states in these systems. We also discuss the possible amplification 
of electron-phonon coupling strength in the presence of the AF order in ThFeAsN.
Finally, we examine the influence of anion As height on the low energy electronic structure especially
the Fermi surface of ThFeAsN. In Sec. V, the conclusions drawn from  this work are summarized.

\section{CRYSTAL STRUCTURE AND COMPUTATIONAL METHODS}
\begin{figure}
\centering
\includegraphics [width=6cm]{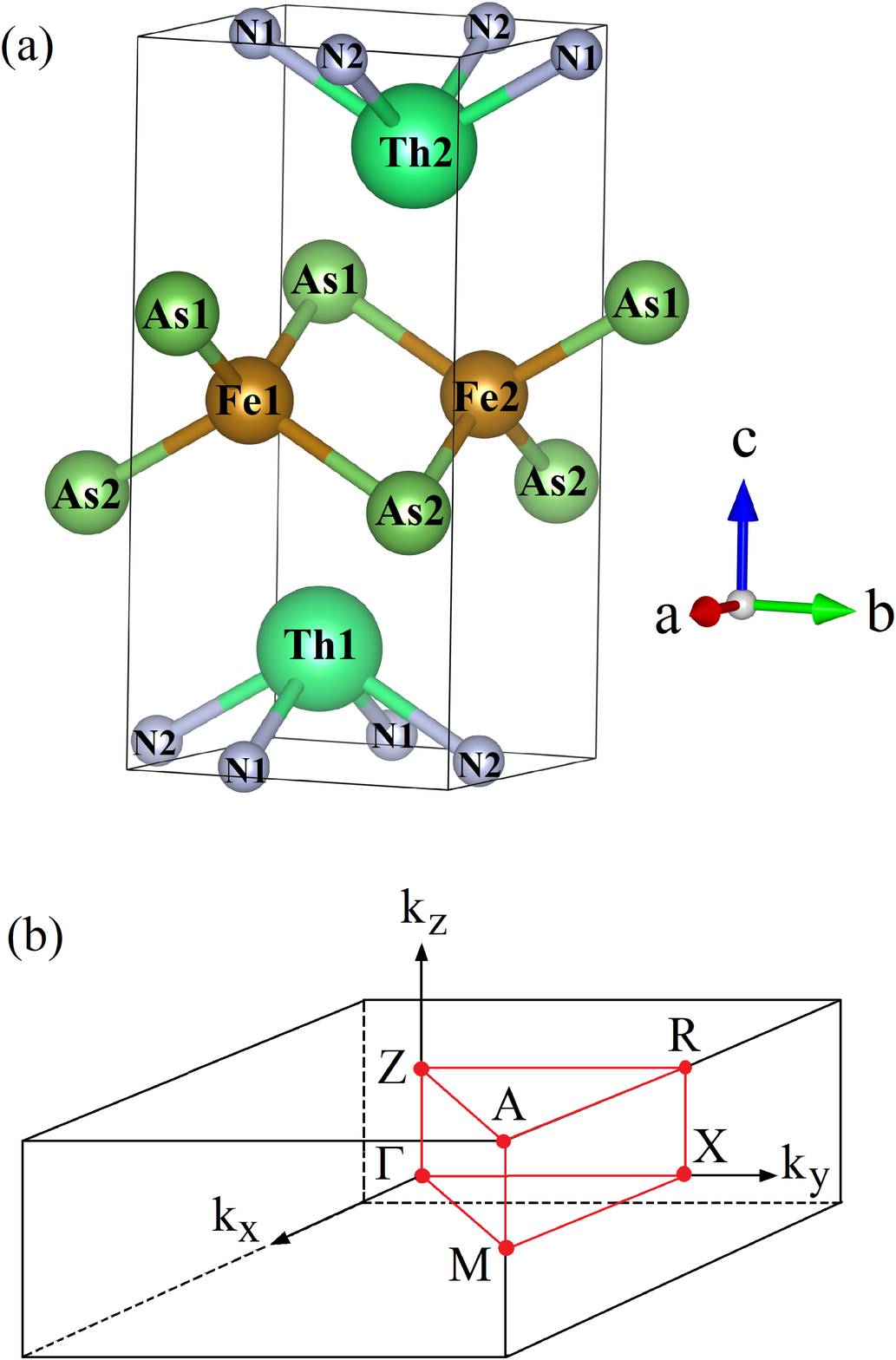}
\caption{(a) Crystal structure and (b) Brillouin zone of ThFeAsN. ThNiAsN and ThCoAsN 
possess the same crystal structure with Fe atoms being replaced by Ni and Co atoms, respectively.}
\label{CS}
\end{figure}

All three considered ThXAsN (X = Fe, Co, Ni) compounds have the same tetragonal symmetry with space group $P4/nmm$ (No. 129) 
and contain two formulas units (f.u.) per unit cell, as illustrated in Fig. \ref{CS}.
Our first principles structural optimization as well as electronic structure, mechanical and lattice dynamical property calculations
are based on the density functional theory with the generalized-gradient approximation (GGA) to the exchange-correlation potential~\cite{Perdew1996}.
For the structural optimization, electronic structure and elastic constant calculations, tbe accurate projector-augmented wave
(PAW) method, as implemented in the Vienna \textit{ab-initio} simulation package (VASP)~\cite{Kresse1993,Bloch1994,Kresse1996}, is used.
The wave functions are expanded in the plane waves basis with a large energy cut-off of $600$ eV. 
The Brillouin zone integration is performed using a $\Gamma$-cantered $10\times10\times5$ Monkhorst-Pack grid. 
For the Fermi surface calculations, a denser $k$-point grid of $20\times20\times20$ is chosen.
In the structural optimizations, the atomic positions are relaxed until the atomic forces are less than $0.0001$ eV/\AA.
In the self consistent total energy and electronic structure calculations, an energy convergence of $10^{-6}$ eV is adopted, 

The available experimental lattice constants of ThFeAsN~\cite{Mao2017} and ThNiAsN~\cite{Wang2017} 
(see Table I) are used, whereas the atomic positions
are determined theoretically by our structural optimization, as described above. 
Since ThCoAsN has not been prepared experimentally, both lattice constants and atomic positions are determined theoretically.
In all these structural optimization calculations, the nonmagnetic (NM) state is assumed unless stated otherwise.
The lattice constants and atomic positions of all the three systems used in the present calculations
of the elastic constant, electronic structure and lattice dynamics, are listed in Table \ref{table3}.  

\begin{table}[h!]
\begin{center}
\caption{Lattice constants ($a,c$) as well as theoretical atomic fractional $z$ coordinates 
and anion As height ($h_{As}$)  of ThXAsN (X = Fe, Co, Ni). For ThFeAsN and ThNiAsN,
the experimental lattice constants are used. For ThCoAsN, 
lattice constants are also determined theoretically (see Sec. II).}
    \vspace{2mm}
    \label{table3}
    \begin{tabular}{lccccc} 
     \hline \hline
              & $a$ & $c$ & $z_{As}$ & $z_{Th}$ & $h_{As}$ (\AA) \\ \hline
      ThFeAsN & 4.0414$^a$  & 8.5152$^a$ & 0.6388 & 0.1382  & 1.18\\
      ThCoAsN & 4.0570 & 8.3519 & 0.6404 & 0.1377  & 1.17\\ 
      ThNiAsN & 4.0804$^b$ & 7.9888$^b$ & 0.6405 & 0.1396  & 1.12\\ \hline \hline
      \end{tabular}
\end{center}
     $^a$ Reference ~\cite{Mao2017}; $^b$ Reference ~\cite{Wang2017}.
\end{table}

The elastic constants of the ThXAsN compounds are determined by using the linear-response stress-strain method,
as implemented in the VASP code \cite{Page2002}. Under a small strain ($\epsilon_{kl}$), according to the Hooke’s law,
the corresponding stress ($\sigma_{ij}$) can be written as $\sigma_{ij}=C_{ijkl}\epsilon_{kl}$, where $C_{ijkl}$ 
is the stiffness tensor, which consists of the elastic constants of the crystal. The total number of independent elastic constants
depends on the crystal symmetry. For example, a cubic crystal has only three independent elastic constants of
$C_{11}$, $C_{12}$ and $C_{44}$.~\cite{Guo2000,Babu2019} On the other hand, the present ThXAsN systems with the tetragonal symmetry 
have six independent elastic constants, namely, $C_{11}$, $C_{12}$, $C_{13}$, $C_{33}$, $C_{44}$ and $C_{66}$.~\cite{Guo2000,Babu2019,Shein2009} 

Within  the Voigt-Reuss-Hill (VRH) approximation~\cite{Hill1952},
the macroscopic bulk moduli ($B$) and shear moduli ($G$) of a polycrystalline material 
can be estimated using the elastic constants of the single crystal mentioned in the preceeding paragraph,
as $B=(B_v+B_R)/2$ and $G=(G_v+G_R)/2$.
Here
$B_v=\frac{1}{9}\{2(C_{11}+C_{12})+C_{33}+4C_{13}\}$ and 
$G_v=\frac{1}{15}\{2C_{11}-1C_{12}+C_{33}-2C_{13}+6C_{44}+3C_{66}\}$ are the minimum bulk moduli and shear moduli 
of the single crystal, respectively.
$B_R=\{C_{33}(C_{11}+C_{12})-2C_{13}^2\}/\{C_{11}+C_{12}+2C_{33}-4C_{13}\}$ and 
$G_R=15/\{18B_v/C^2+6/(C_{11}-C_{12})+6/C_{44}+3/C_{66}\}$ are the maximum bulk moduli and shear moduli
of the single crystal, respectively.
Furthermore, the Young's modulus ($Y$) and Poisson's ratio ($\nu$) of the polycrystalline material can be written 
in terms of $B$ and $G$ as $Y=9BG/(3B + G)$ and $\nu=(3B-2G)/2(3B+G)$.
The averaged sound wave velocity of the polycrystalline material can be written as~\cite{Anderson1963}
$\overline{v}_m=[\frac{1}{3}(\frac{2}{\overline{v}_t^3}+\frac{1}{\overline{v}_l^3})]^{-1/3}$
where $\overline{v}_l$ and $\overline{v}_t$ are the averaged longitudinal and transverse elastic wave
velocities, respectively, which are related to $B$ and $G$ as
$\overline{v}_l=(\frac{B+4G/3}{\rho})^{1/2}$
and
$\overline{v}_t=(\frac{G}{\rho})^{1/2}$.
Finally, the Debye temperature of the polycrystalline material can be estimated from the average sound wave velocity via~\cite{Anderson1963}:
\begin{equation}
\Theta_D=\frac{h}{k_B}[\frac{3n}{4\pi}(\frac{N_A\rho}{M})]^{1/3}\overline{v}_m
\end{equation}
where $h$ is the Plank's constant; $k_B$ is the Boltzmann's constant; $N_A$ is
the Avogadro's number; $\rho$ is the density; $M$ is the molecular weight
and $n$ is the number of atoms per unit cell. 

All the phonon and electron-phonon coupling calculations are performed using the plane wave pseudopotential method, as implemented in
the Quantum ESPRESSO package~\cite{Giannozzi2009}.  All the pseudopotentials are taken from the Pslibrary~\cite{PSlibrary}.
The phonon dispersions and electron-phonon coupling  are calculated based on the first-principles 
density functional perturbation theory (DFPT)~\cite{Baroni2001}. We choose the plane wave energy cut-off as 55 Ry and the
charge density cut-off as 220 Ry. A Gaussian broadening method with the Gaussian width of
0.02 Ry is used in the Brillouin zone integration. A $k$-mesh of $12\times12\times6$ for electrons 
and a $q$-mesh of $6\times6\times3$ for phonon are used for all three systems.
The electron-phonon coupling constant ($\lambda$) can be evaluated using the relation \cite{McMillan1968,Allen1975}:
\begin{equation}
\lambda=2\int\frac{\alpha^2F(\omega)}{\omega}d\omega,
\end{equation}
where the Eliashberg spectral function 
\begin{equation}
\alpha^2F(\omega)=\frac{1}{2\pi N(E_F)}\sum\frac{\gamma_{qj}}{\omega_{qj}}\delta(\hslash\omega-\hslash\omega_{qj}).
\end{equation}
Here $N(E_F)$ is the electronic density of states at the Fermi level ($E_F$ ), 
$\gamma_{qj}$ is the phonon linewidth due to electron-phonon scattering, $\omega_{qj}$ is the phonon frequency with wave vector $q$
and branch index $j$. Finally, the phonon-mediated superconducting transition temperature in Kelvin, can be evaluated using the Allen
and Dynes-modified McMillan’s formula \cite{McMillan1968,Allen1975}:
\begin{equation}
T_c=\frac{\omega_{log}}{1.2}exp[\frac{-1.04(1+\lambda)}{\lambda-\mu^*(1+0.62\lambda)}]
\end{equation}
where $\omega_{log}$ is the logarithmic average phonon frequency, and $\mu^*$ is the averaged screened electron-electron interaction.

\section{Nonmagnetic state properties}
In this section, we present the calculated physical properties of ThXAsN (X = Fe, Co, Ni) in
the nonmagnetic state.

\subsection{Elastic constants and mechanical stability}
All calculated elastic constants are presented in Table \ref{table1}.
They all are positive (see Table \ref{table1}) 
and obey the well known Born criterion of mechanical stability \cite{Born}.
Table \ref{table1} indicates that the $C_{33}$ values of all the three systems are well above the limit 
of the elastic stability, \textit{i.e.}, $C_{33}   >  C^* =2C^2_{13} /(C_{11}+C_{12})$~\cite{Kittle}.
For all the three systems, we have $B > G$, which indicates that the shear modulus $G$ is the parameter
that dictates the mechanical stability of the systems under consideration. 
Interestingly, Table II indicates that our calculated values of the elastic constants and elastic modulus 
are similar to the theoretical values of LaFeAsO \cite{Shein2009}. 
Although LaFeAsO and ThFeAsN have a very similar lateral lattice constant ($a=4.035$ \AA$ $ 
for LaFeAsO~\cite{Boeri2008}), the vertical lattice constant $c$ of ThFeAsN is much smaller than 
that of LaFeAsO ($c=8.741$ \AA$ $~\cite{Boeri2008}). 
This pronounced difference in $c$ would cause some internal uniaxial pressure in ThFeAsN along the $c$-axis. 
In spite of this pronounced structural difference, mechanical properties and elastic constants
of these two compounds are very similar (Table II).

However, our calculated bulk modulus of ThFeAsN is lower than 
the bulk modulus estimated theoretically by Albedah et al. \cite{Albedah2017}. 
It turns out that the bulk modulus of these systems ($\sim$100 GPa) are smaller than 
that of other classes of high $T_c$ SCs such as MgB$_2$ ($\sim$122-161 GPa)~\cite{Osorio-Guillen2002} 
and YBa$_2$Cu$_3$O$_7$ ($\sim$200 GPa)~\cite{Lei1993}. Thus, in
comparison to these high $T_c$ SCs, ThXAsN (X=Fe,Co,Ni) are relatively soft materials with high compressibilities.
Young's modulus determine the stiffness of materials. Therefore, ThNiAsN has the highest and ThCoAsN has the lowest stiffness.
A material is considered to be brittle if it satisfies the well known Pough's criterion $B/G <1.75$. Our results indicate that 
ThFeAsN lies in the boundary of brittle and plastic states as that of the LaFeAsO compound. On the contrary, ThNiAsN and ThCoAsN systems clearly show plastic 
behaviour. The values of Poisson's ratios indicate the ionic nature of bonding in these three compounds.

\begin{table}[h!]
  \begin{center}
\caption{Calculated elastic constants ($C_{ij}$), bulk modulus
($B_H$), shear modulus ($G_H$) and Young’s modulus ($Y$) in GPa 
as well as Debye temperature ($\Theta_D$), Poisson's ratio ($\nu$) and compressibility ($\beta_{H}$ in GPa$^{-1}$) 
of ThXAsN (X = Fe, Co, Ni). 
For comparison, the properties of LaFeAsO (a parent compound 
of the 1111 Fe-based superconductor family) are also listed here.}
    \vspace{2mm}
    \label{table1}
    \begin{tabular}{lcccc} 
     \hline \hline
             & ThFeAsN & ThCoAsN & ThNiAsN & LaFeAsO*\\
    \hline
     $C_{11}$ & 211 & 211 & 235 & 192 \\
     $C_{12}$ & 74 & 64 & 76  & 56\\
     $C_{13}$ & 68 & 72 & 85 & 62\\
     $C_{33}$ & 105 & 109 & 120 & 145\\
     $C_{44}$ & 95 & 57 & 74 & 44 \\
     $C_{66}$ & 39 & 44 & 56 & 78\\
     $B_H$ & 99 & 101 & 114 & 99\\
     $G_H$ & 61 & 51  & 60  & 57 \\
      $Y$ & 151 & 130 & 153 & 142 \\
      $\nu$ & 0.24 & 0.28 & 0.27 & 0.25\\
      $\beta_{H}$ & 0.0101 & 0.0099 & 0.0087 & 0.0102\\
      $\Theta_D$ (K) & 417 & 381 & 385 & 441 \\
      \hline \hline
      \end{tabular}
  \end{center}
  *Reference \cite{Shein2009}
\end{table}

\subsection{Plasma frequency and effective dimensionality}
Electronic structure calculations can provide important
information about the ‘effective dimensionality’ of a system through various quantities such as 
dispersion-less (flat) energy bands along certain symmetry lines, 
and van Hove singularities in density of states. A simple quantitative estimate 
of the ‘effective dimensionality’ can be acquired by calculating plasma frequencies
along the principle axes of a crystal \cite{Kasinathan2009}. 
The plasma frequency tensor $\omega^2_{P(\alpha \beta)}$ can be calculated from first principles as an integral over the Fermi surface using the relation \cite{Setten2009}:
\begin{equation*}
{\omega }^2_{P(\alpha \beta )}=-\frac{4\pi e^2}{V{\textrm{$\hslash$}}^2}\sum_{n,\boldsymbol{k}}
{2g_{\boldsymbol{k}}}\frac{\partial f\left({\epsilon }_{n\boldsymbol{k}}\right)}{\partial \epsilon }
\left({\boldsymbol{e}}_{\alpha }\frac{\partial {\epsilon }_{n\boldsymbol{k}}}{\partial \boldsymbol{k}}\right)\left
({\boldsymbol{e}}_{\boldsymbol{\beta }}\frac{\partial {\epsilon }_{n\boldsymbol{k}}}{\partial \boldsymbol{k}}\right)
\end{equation*}
where $g_k$ is the weight factor of the \textbf{k} point, $f(\epsilon_{n\textbf{k}})$ is the Fermi-Dirac function, 
and \textbf{$e_{\alpha}$} and \textbf{$e_{\beta}$} are the unit vectors in the $\alpha$ and $\beta$ directions. 
respectively. $V$ is the unit cell volume, $e$ is the electron charge, 
$\hslash$ is the reduced Planck constant and $\epsilon_{n\textbf{k}}$ are the energy eigenvalues.

\begin{table}[h!]
  \begin{center}
\caption{Ratio of the in-plane plasma frequency to the out-of-plane plasma frequency ($\omega^a_P/\omega^c_P$),
$c/a$ ratio, ratio of out-of-plane and in-plane Fe(Ni/Co)-Fe(Ni/Co) distances ($d^c_{X-X}/d^a_{X-X}$) and superconducting
transition temperature ($T_c$) of some Fe-based SCs along with ThXAsN (X = Fe, Co, Ni).}
    \vspace{2mm}
    \label{table2}
    \begin{tabular}{lcccc} 
     \hline \hline
    {System} & \textbf{$\omega^a_P/\omega^c_P$} & \textbf{c/a} & \textbf{$d^c_{X-X}/d^a_{X-X}$} & Expt. \textbf{$T^{max}_c$}(K)\\
    \hline
      SrFeAsF$^a$ & 19.892 & 2.2426 & 3.1715 & 56$^b$\\
      LaFeAsO$^a$ & 8.9467 & 2.1656     & 3.0626 & 52$^c$\\
      FeSe$^a$ & 4.1119 & 1.4656 & 2.0727 & 8$^d$\\
      LiFeAs$^a$ & 3.2181 & 1.6785 & 2.3738 & 18$^e$\\
      BaFe$_2$As$_2$$^a$ & 3.2926 &     3.285 & 2.3228 & 38$^f$\\
      ThFeAsN & 3.0621 & 2.1069 & 2.9797 & 30$^g$\\
      ThCoAsN & 6.3682 & 2.06 & 2.9144 & -\\
      ThNiAsN & 7.0934 & 1.957 & 2.7683 & 4.3$^h$\\
      \hline \hline
      \end{tabular}
  \end{center}
  $^a$Reference \cite{Kasinathan2009}\\
  $^b$Reference \cite{Wu2009} (doped compound); $^c$Reference \cite{Takahashi2015} (doped compound); $^d$Reference \cite{Hsu2008};
  $^e$Reference \cite{Tapp2008}; $^f$Reference \cite{Rotter2008} (doped compound); $^g$Reference \cite{Wang2016};
  $^h$Reference \cite{Wang2017}.
\end{table}

Here we evaluate the plasma frequencies of all ThXAsN (X = Fe, Co, Ni) along the crystalline axes,
and compare our results with some other Fe-based SCs~\cite{Kasinathan2009}, 
including LaFeAsO, one of the parent compounds of the 1111 family.
In Table \ref{table2},  we depict the calculated ratio of the in-plane plasma frequency $\omega^a_P$ 
to the  out-of-plane plasma frequency $\omega^c_P$. Also in Table \ref{table2}, 
we present some important structural parameters as well as superconducting transition temperature for these SCs. 

It is evident from Table \ref{table2} that the plasma frequency ratio 
of ThFeAsN is significantly smaller than that of LaFeAsO. This implies that ThFeAsN is more 3D-like
 which in turn make it least anisotropic among all the compounds listed in Table \ref{table2}. 
Moreover, the value of $c/a$ ratio as well as $d^c_{Fe-Fe}/d^a_{Fe-Fe}$ 
are also smaller than that of LaFeAsO. More 3D-like electronic structure goes against the Fermi surface nesting, 
and this may be the reason of not having the SDW ordering which was observed in LaFeAsO. 
On the other hand, the plasma frequency ratios in ThCoAsN and ThNiAsN are higher than that of ThFeAsN. 
This also indicates that ThCoAsN and ThNiAsN possess a 
more 2D-like electronic structure as compared to ThFeAsN.
The ratio of the shortest interlayer to the shortest intra-layer Fe(Ni/Co)-Fe(Ni/Co) distance 
is $c/\sqrt{2}a$ for the 122 systems and is $\sqrt{2}c/a$ for the others including the three systems studied here. 
However, ThFeAsN violates the general trend of proportionality of 
$\omega^a_P/\omega^c_P$  to that of $d^c_{Fe-Fe}/d^a_{Fe-Fe}$ for the Fe-pnictide SCs.

\begin{figure}[h!]
\centering
\includegraphics [width=7.5cm]{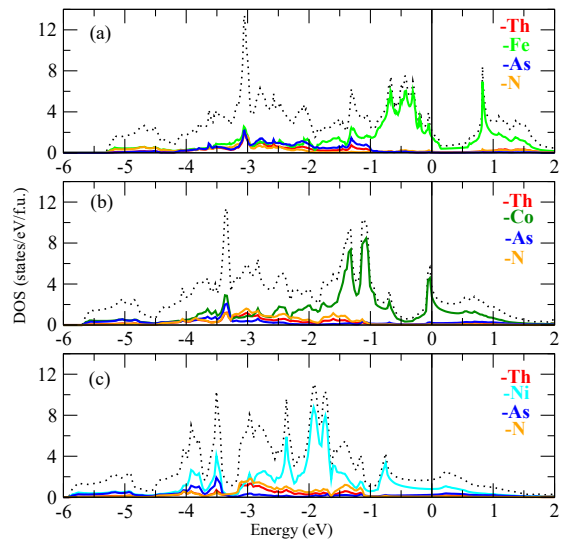}
\caption{Calculated total (dotted line) and atom-projected density of states of (a) ThFeAsN, (b) ThCoAsN and (c) ThNiAsN.
The Fermi energy is at 0 eV.}
\label{LDOS}
\end{figure}

\begin{figure}[h!]
   \centering
   \includegraphics [width=8.5cm]{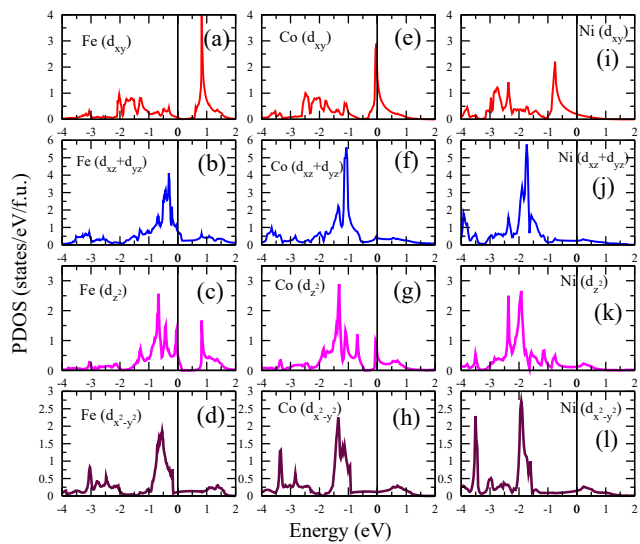}
\caption{Calculated transition metal $d$ orbital-projected
density of states (PDOS) of (a-d) ThFeAsN, (e-h) ThCoAsN, and (i-l) ThNiAsN.}
\label{PDOS}
\end{figure}

\begin{figure*}
   \centering
   \includegraphics [width=12.0cm]{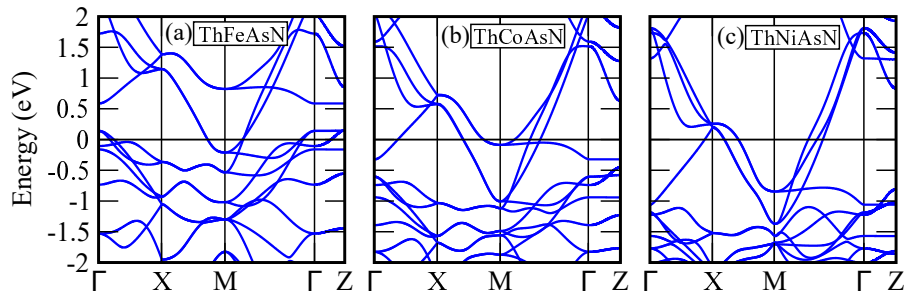}
   \caption{Calculated band structures of (a) ThFeAsN, (b) ThCoAsN and (c) ThNiAsN.}
   \label{allbs}
\end{figure*}

\subsection{Electronic structure}
Low energy electronic structure plays a key role in superconductivity.
In this subsection, therefore, we present the calculated electronic structures of ThXAsN (X = Fe, Co, Ni).
First, we display the total and atom-projected density of states (DOS) for ThFeAsN, ThCoAsN and ThNiAsN
in Fig. \ref{LDOS}. We notice that overall, the DOS spectra for these three compounds are rather similar
except that the Fermi levels ($E_F$) for the three compounds are located at the different positions.
This may be expected since they are isostructural.
In ThFeAsN, the DOS near $E_F$ is dominated by transition metal (X) (Fe in this case) $d$-orbitals, 
which form two peaks, namely, one broad peak just below $E_F$ and one sharp peak located at $\sim$0.8 eV above $E_F$ [see Fig. 2(a)].
There is a pseudo gap between the two DOS peaks and the Fermi level is located at the lower edge of the pseudo gap.
This gives rise to a moderate DOS at $E_F$ of 2.3 states/eV/f.u. for ThFeAsN.
As one moves from ThFeAsN to ThCoAsN, ThCoAsN has an additional valence electron and thus 
the Fermi level is raised to sit nearly right on the upper peak [see Fig. 2(b)].
This results in a large DOS at $E_F$ of 3.6 states/eV/f.u. for ThCoAsN. 
In ThCoAsN, thus, a sharp peak in the DOS (mainly coming from Co atoms) at the Fermi level is observed.
A large DOS at $E_F$ may be important for superconductivity as it increases the possibility of electron-electron pairing.
Unfortunately, a large DOS at $E_F$ would also cause the Stoner instability~\cite{Stoner1939,Janak1977}, 
leading to the ferromagnetism in the Co system (see the next section).
As one moves onto ThNiAsN, $E_F$ is further raised to the rather flat upper part of the $d$ dominated
band above the sharp peak [see Fig. 2(c)] because ThNiAsN has one more electron than ThCoAsN.  
Figure 2(c) shows that in this energy range, there are significant contributions from anion $p$ orbitals,
indicating a significant hybridization among the Ni $d$ and anion $p$ orbitals.
This leads to the formation of rather dispersive bands across the Fermi level [see Fig. 4(c)],
and thus results in a lowest DOS at $E_F$ (1.7 states/eV/f.u.) for ThNiAsN among the three compounds.
We notice that our calculated DOS for ThFeAsN and ThNiAsN agree well with previous experiments as well
as theoretical calculations \cite{2Wang2016,Shiroka2017,Singh2016,Albedah2017}.

\begin{figure}[h!]
   \centering
   \includegraphics [width=8.5cm]{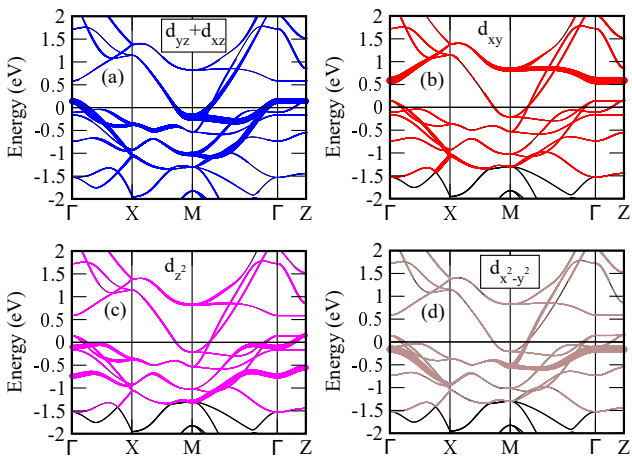}
\caption{Calculated Fe (a) $d_{yz}+d_{xz}$ (blue), (b) $d_{xy}$ (red), (c) $d_{z^2}$ (magenta) 
and (d) $d_{x^2-y^2}$ (brown) orbital-projected band structures of ThFeAsN.}
\label{Feorb}
\end{figure}

\begin{figure}[h!]
   \centering
   \includegraphics [width=8.5cm]{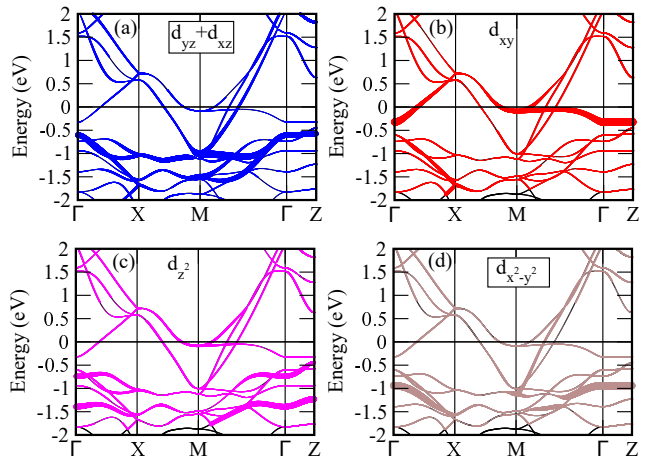}
\caption{Calculated Co (a) $d_{yz}+d_{xz}$ (blue), (b) $d_{xy}$ (red), (c) $d_{z^2}$ (magenta) 
and (d) $d_{x^2-y^2}$ (brown) orbital-projected band structures of ThCoAsN.}
\label{Coorb}
\end{figure}

In Fig. \ref{PDOS}, we show the X $d$ orbital-projected DOS of ThXAsN (X = Fe, Co, Ni). 
In the Fe system, $d_{yz}$, $d_{xz}$ and $d_{z^2}$ orbitals dominate the DOS in the vicinity of   
the Fermi level whereas in the Co system, the DOS near $E_F$ is mainly derived from the $d_{xy}$ orbital. 
In fact, the sharp peak in the DOS for Co $d_{xy}$ orbital is observed at the Fermi level.
Lastly in the Ni system, no particular dominance of any particular orbital is observed. 
Overall, ThFeAsN and ThCoAsN show more multi-orbital characteristics near $E_F$ than that of ThNiAsN. 
Pseudo-gap like regions in the DOS of ThFeAsN and ThCoAsN are mainly due to the $d_{z^2}$ orbital
which is extended along the $c$-axis. 

\begin{figure}[h!]
   \centering
   \includegraphics [width=8.5cm]{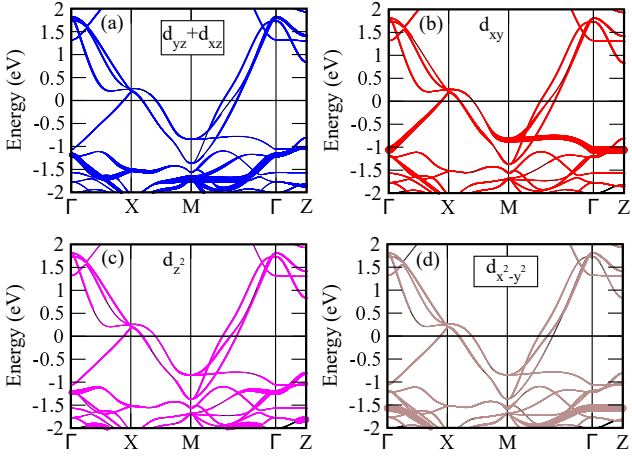}
   \caption{Calculated Ni (a) $d_{yz}+d_{xz}$ (blue), (b) $d_{xy}$ (red), (c) $d_{z^2}$ (magenta) 
    and (d) $d_{x^2-y^2}$ (brown) orbital-projected band structures of ThNiAsN.}
   \label{Niorb}
\end{figure}

\begin{figure}[h!]
   \centering
   \includegraphics [width=8.5cm]{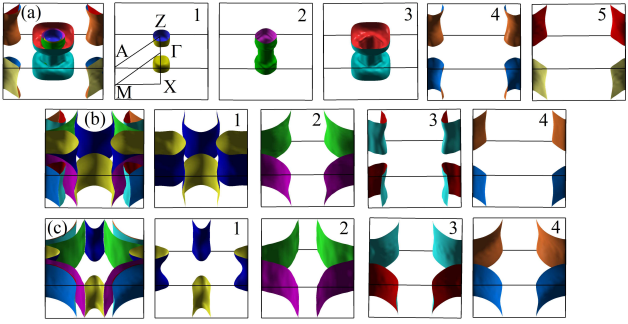}
\caption{Calculated Fermi surface (FS) of (a) ThFeAsN, (b) ThCoAsN and (c) ThNiAsN. In ThFeAsN, 
there are 3 hole FSs (panels 1-3) at the $\Gamma$ point and 2 electron FSs (panels 4 and 5) at the M point. 
In ThCoAsN and ThNiAsN, there are 1 hole FS (panel 1) at the X point and 3 electron FSs (panels 2-4) at the M point.}
   \label{FS}
\end{figure}

Next we present the calculated band structures of ThXAsN (X = Fe, Co, Ni) in Fig. \ref{allbs}. 
We also display the orbital-projected band structures of ThFeAsN, ThCoAsN and ThNiAsN  
in Figs. \ref{Feorb}, \ref{Coorb} and \ref{Niorb}, respectively.
Electronic band dispersions near the Fermi level for the Fe, Co and Ni systems are quite different from each other. 
In particular, the band structure of ThFeAsN is significantly different from the other two systems. 
Band structure of ThFeAsN is quite similar to that of F-doped LaFeAsO~\cite{Prando2015}. 
Interestingly, band structure of ThCoAsN possesses a Co $d_{xy}$ orbital dominated flat band just below $E_F$ 
along the high symmetry $\Gamma$-M line [see Fig. 4(b) and Fig. 6]. 
Orbital characters of the bands near $E_F$ are also different for all three systems. 
In ThFeAsN, a rather flat band near $E_F$ coming from Fe $d_{z^2}$ orbital is observed around the $\Gamma$ point (Fig. 5). 
On the other hand, this particular feature is missing in the band structure of ThNiAsN. Flat bands indicate a large number of
electronic states.  Therefore, the presence of flat bands near $E_F$ 
would enhance the possibility of electron pairing (formation of Cooper pairs), irrespective of superconducting pairing mechanism.

Calculated Fermi surfaces (FSs) of the ThXAsN compounds are depicted in Fig. \ref{FS}. 
For the Fe system, the FS consists of 3 FS pockets [panels 1-3 in Fig. 8(a)] at the $\Gamma$ point 
and 2 electron FS sheets [panels 4 and 5 in Fig. 8(a)] at the M ($\pi$, $\pi$, 0) point in the Brillouin zone. 
This Fermi surface topology of ThFeAsN is quite similar to that of other Fe-based SCs 
such as LaFeAsO \cite{Sen,Singh2008}. One of the three hole FS pockets (panel 2) at the $\Gamma$ point 
is nearly two dimensional and thus looks like a circular cylinder, 
whereas the 2 electron FS sheets (panels 4 and 5) at the zone corners (the M points) are more 3D like. 
For the Co and Ni systems, in contrast, there are 3 electron FS sheets (panels 2-4) around the M point 
and 1 hole FS (panel 1) at the X point. There is no FS sheet at the $\Gamma$ point in these two systems. 
Since the appearance of FSs is at different points for ThFeAsN and ThCo(Ni)AsN, interband nesting 
vectors are different for these systems.
These significant differences may affect the intraband and interband nesting properties of 
these system, which could play a role in the possible formation of spin and charge density waves as well as superconductivity 
in this type of SCs. The absence of hole FS pockets at the centre of the Brillouin zone in ThCoAsN and ThNiAsN 
does not help interband FS nesting, and hence these two systems are not likely to show FS nesting-driven 
spin/charge density wave ordering. 

\begin{table}
\begin{center}
\caption{Calculated electronic charges by the Bader analysis, and also ideal ionic charges for the atomic species in
ThXAsN (X = Fe, Co, Ni).}
    \vspace{2mm}
    \label{Bader}
    \begin{tabular}{ccc} 
     \hline
    \hline
    Species & Bader analysis & Purely ionic \\
    \hline
    \multicolumn{3}{c}{(a) ThFeAsN}\\
    \hline
     Th & 9.77 & 8 (Th$^{4+}$) \\
      Fe & 7.88 & 6 (Fe$^{2+}$) \\
                   As & 5.73 & 8 (As$^{3-}$) \\
                    N & 6.61 & 8 (N$^{3-}$) \\
                     ThN & 16.38 & 16 (ThN$^+$) \\
                       FeAs & 13.61 & 14 (FeAs$^-$)\\
      \hline 
      \multicolumn{3}{c}{(b) ThCoAsN}\\ 
          \hline
          Th & 10.22 & 8 (Th$^{4+}$) \\
            Co & 8.99 & 7 (Co$^{2+}$) \\
             As & 5.49 & 8 (As$^{3-}$) \\
              N & 6.29 & 8 (N$^{3-}$) \\
               ThN & 16.51 & 16 (ThN$^+$) \\
                 CoAs & 14.48 & 15 (CoAs$^-$)\\
     \hline 
     \multicolumn{3}{c}{(c) ThNiAsN}\\ 
         \hline
          Th & 9.81 & 8 (Th$^{4+}$) \\
           Ni & 10.04 & 8 (Ni$^{2+}$) \\
            As & 5.56 & 8 (As$^{3-}$) \\
             N & 6.59 & 8 (N$^{3-}$) \\
              ThN & 16.40 & 16 (ThN$^+$) \\
               NiAs & 15.60 & 16 (NiAs$^-$)\\
               \hline\hline
      \end{tabular}
      \end{center}
  \end{table}

We also perform a Bader analysis \cite{Bader} on the charge densities in ThXAsN (X = Fe, Co, Ni),
and the results are listed in Table \ref{Bader}.
In ThFeAsN, a loss of 0.38 electrons in the [FeAs]$^-$ layer is observed (see Table \ref{Bader}),
which is consistent with the previous theoretical study~\cite{2Wang2016}. 
In ThCoAsN and ThNiAsN, we find a loss of 0.51 and 0.40 electrons in the [Co/NiAs]$^-$ layer, respectively.
In ThCoAsN, the interlayer charge transfer is larger as compared to the other two systems.
More charge transfer between the adjacent layers indicates the enhancement of the ionic character in the interlayer bonding. 
As a result, the interlayer bonds which are responsible for the common stability of the crystal are weakest 
in ThFeAsN and strongest for ThCoAsN. Thus, ThFeAsN is softer than ThCoAsN, 
and the softness of ThNiAsN lies between them.

\subsection{Phonon dispersion and electron-phonon coupling}
\begin{figure}[h!]
   \centering
   \includegraphics [width=7.0cm]{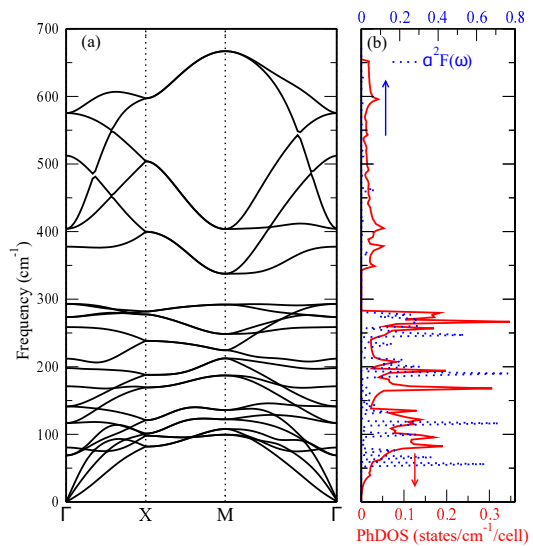}
\caption{(a) Phonon dispersion relations, (b) Eliashberg spectral function [$\alpha^2F(\omega)$] (blue dotted line)     
and phonon density of states (PhDOS) (red line) of ThFeAsN.}
\label{Fe}
\end{figure}

\begin{figure}[h!]
   \centering
   \includegraphics [width=7.0cm]{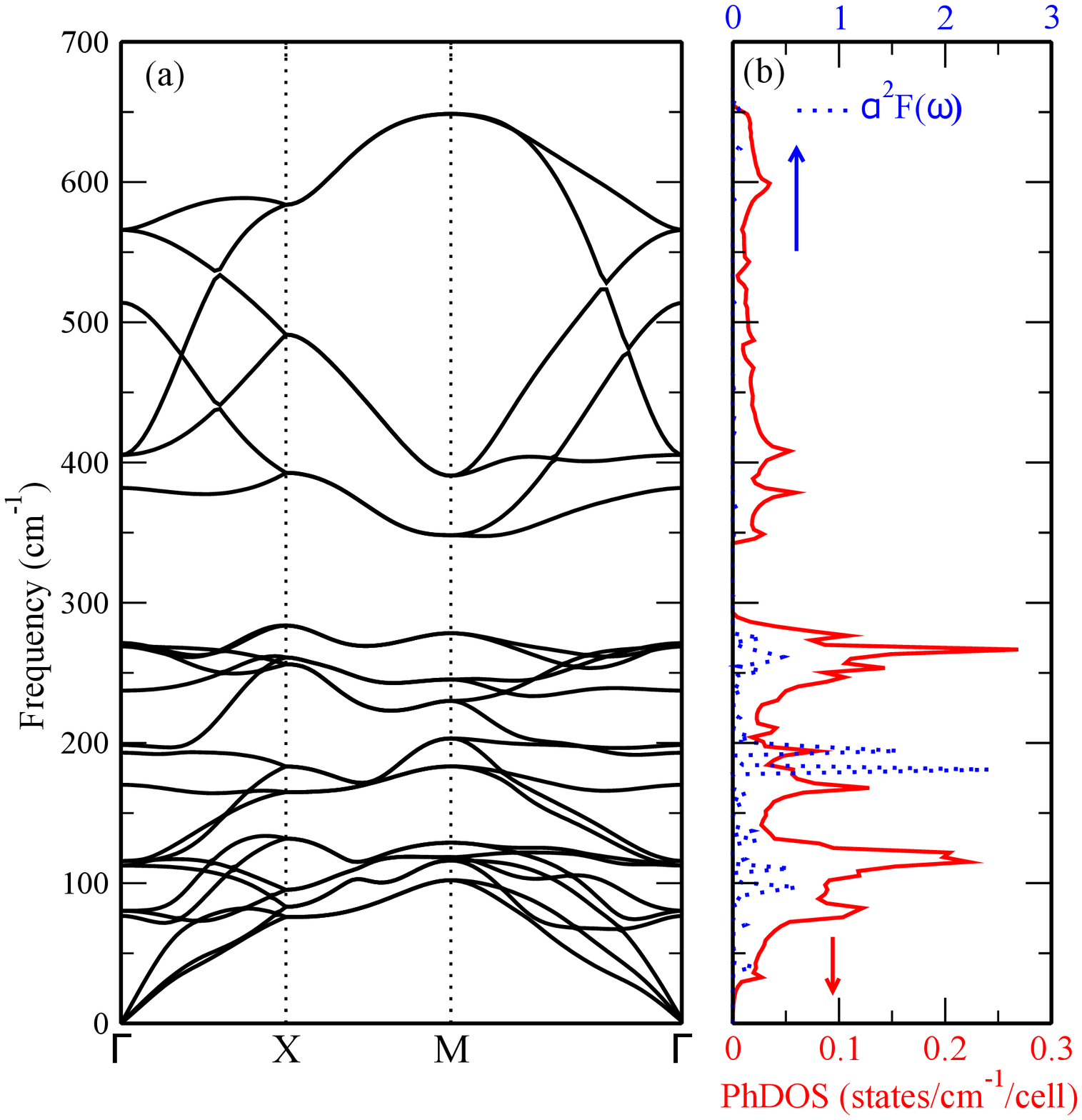}
\caption{(a) Phonon dispersion relations, (b) Eliashberg spectral function [$\alpha^2F(\omega)$] (blue dotted line)
and phonon density of states (PhDOS) (red line) of ThCoAsN.}
\label{Co}
\end{figure}

\begin{figure}[h!]
   \centering
   \includegraphics [width=7.0cm]{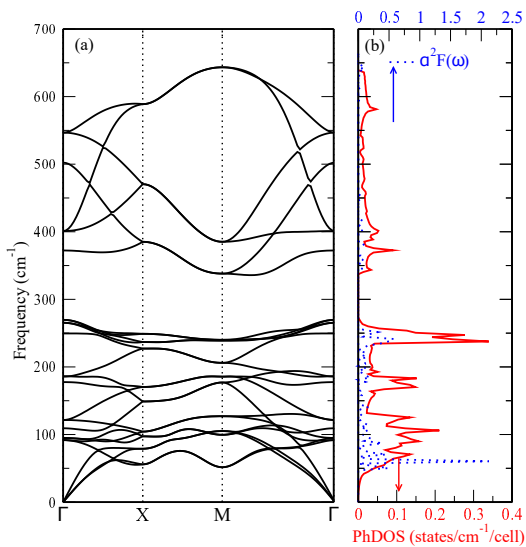}
\caption{ (a) Phonon dispersion relations, (b) Eliashberg spectral function [$\alpha^2F(\omega)$] (blue dotted line)
and phonon density of states (PhDOS) (red line) of ThNiAsN.}
\label{Ni}
\end{figure}

Now let us examine the calculated lattice dynamical properties of ThXAsN (X = Fe, Co, Ni). 
In Figs. \ref{Fe}(a) and \ref{Co})(a) and \ref{Ni}(a), we present the calculated phonon dispersion relations 
of ThFeAsN, ThCoAsN and ThNiAsN, respectively.
Since all three systems have 8 atoms per unit cell, there are 24 branches with three acoustic and 21 optical phonon branches. 
First of all, it is clear from these phonon dispersion relations that there are no 
imaginary frequencies, thus indicating that these systems are dynamically stable. 
Second, there exists a clear gap in the phonon dispersion relations in all the three systems (see Figs. \ref{Fe}, \ref{Co} and \ref{Ni}). 
Interestingly, the gap in the Fe system is smaller than that of the Co and Ni systems. 
This difference is due to the differences in the atomic mass among the Fe, Co and Ni atoms. 
The six high lying optical branches above the gap are mainly due to the lighter N and As atoms,
thus being almost the same in all three systems. 
On the other hand, the 18 lower frequency branches including three acoustic modes below the gap, come predominantly from the Th and 
transition metal atoms. Consequently, the band width of these lower frequency branches becomes narrower as
Fe is placed by heavier Co and Ni, and this gives rise to a slightly larger gap in ThCoAsN (Fig. \ref{Co}) and ThNiAsN (Fig. \ref{Ni})
than in ThFeAsN (Fig. \ref{Fe}).  
We notice that phonon dispersion relations of ThFeAsN are quite different from that of LaFeAsO~\cite{Boeri2008}.
In particular, there is no such gap as mentioned above in LaFeAsO~\cite{Boeri2008}.
We also notice that although the lattice dynamical properties of ThNiAsN have already 
been studied theoretically within the DFPT~\cite{Yang2018},
no study on lattice dynamics of ThFeAsN and ThCoAsN has been reported.
The present phonon dispersion of ThNiAsN agree rather well with that reported in ~\cite{Yang2018}.

In Figs. \ref{Fe}(b), \ref{Co}(b) and \ref{Ni}(b), we display the calculated phonon density of states (phDOS)
as well as Eliashberg spectral functions ($\alpha^2F(\omega)$) of ThFeAsN, ThCoAsN and ThNiAsN, respectively.
We notice that the phDOS in the low frequency manifold is many times larger than
the high lying optical branches above the gap, because there are many narrow branches in the manifold.
The $\alpha^2F(\omega)$ spectral functions, which reflect the electron-phonon coupling strength, roughly follow
the phDOS spectra [see Figs. \ref{Fe}(b), \ref{Co}(b) and \ref{Ni}(b)], i.e., we can regard an $\alpha^2F(\omega)$ 
spectral function as the corresponding phDOS spectrum modified by the electron-phonon coupling matrix elements. 
As a result, the Eliashberg function $\alpha^2F(\omega)$ is dominated by the contribution from the phonon bands
in the low frequency manifold for all the three compounds.
By integrating $\alpha^2F(\omega)/\omega$ over the entire frequency range [see Eq. (2)], one can obtain
the electron-phonon coupling constant ($\lambda$).
It is evident from Figs. \ref{Fe}b, \ref{Co}b and \ref{Ni}b that
the magnitude of the $\alpha^2F(\omega)$ spectrum is highest in the Co system and lowest in the Fe system. 
Therefore, we find a lowest $\lambda$ value for the Fe system and a highest $\lambda$ value for the Co system 
(see Table \ref{table4}). Nevertheless, the calculated $\lambda$ value (0.64) for the Ni system
is only very slighty smaller than the Co system (0.65). 

Finally, given the $\lambda$ value as well as other phonon and electron parameters,
we can use the Allen-Dynes-modified McMillan formula [Eq. (4)] to estimate the superconducting transition temperature ($T_c$).
Here we treat the screened electron-electron repulsion $\mu$ as an empirical parameter and set $\mu = 0.10$~\cite{Yang2018}. 
Our calculated $T_c$ values for all the compounds are listed in Table \ref{table4}, together with the available experimental $T_c$ values.
We notice that our calculated $T_c$ of 3.4 K for ThNiAsN agrees very well with the previous calculation (3.5 K)~\cite{Yang2018} 
and also with the experimental $T_c$ of 4.3 K~\cite{Wang2017}. 
However, our calculated $T_c = 0.05$ K for ThFeAsN is orders of magnitude smaller than
the observed $T_c$ (over 30 K)~\cite{Wang2016}.
This is because the calcuated electron-phonon coupling strength $\lambda$ (0.28) for ThFeAsN 
is much smaller than ThNiAsN (0.64). Clearly, this indicates
that electron-phonon coupling cannot be the primarily mechanism of 
the high $T_c$ superconductivity observed in ThFeAsN.
Interestingly, ThCoAsN is predicted to have the highest $T_c$ 
of 6.4 K (Table \ref{table4}). We notice that although the calculated $\lambda$ values
for ThCoAsN and ThNiAsN are nearly equal, the calculated $T_c$ of ThCoAsN is significantly higher
than ThNiAsN. This could be attributed to the fact that the averaged logarithmic phonon frequency
$\omega_{log}$ of ThCoAsN is nearly twice larger than ThNiAsN [see Table \ref{table4} and Eq. (4)].

\begin{table}[h!]
  \begin{center}
    \caption{Calculated electron-phonon coupling constant ($\lambda$), logarithmic average phonon frequency ($\omega_{log}$) , 
    Debye temperature ($\Theta_D$) and superconducting transition temperature $T_c$ of ThXAsN. 
    Experimental $T_c$  values are also listed in brackets.}
    \vspace{2mm}
    \label{table4}
    \begin{tabular}{lcccc} 
     \hline
    {System} & $\lambda$ & $\omega_{log}$ & $N(_{E_F})$ & $T_c$ (K)\\
      &  & (K) & (states/eV/f.u.) & Theory (Expt.)\\
    \hline
     ThFeAsN & 0.28 & 199 & 2.01 & $0.05$ (30$^a$)\\
     ThCoAsN & 0.65 & 221 & 4.22 & 6.4\\
     ThNiAsN & 0.64 & 120 & 1.81 & 3.4 (4.3$^c$)\\
      \hline
      \end{tabular}
  \end{center}
  $^a$Reference \cite{Wang2016}, $^c$Reference \cite{Wang2017}
\end{table}

\section{Discussion}
In this section, we first 
study possible magnetic states in these compounds and the effect of their presence on the band structures.
As reported in the preceding section, our DFPT calculations predict a $T_c$ value for ThFeAsN
which is much too small compared with the observed value (see Table \ref{table4}). 
Therefore, also in this section, we discuss the effect of possible SDW-type AF orders
on the electron-phonon coupling and superconductivity in ThFeAsN. 
We then examine the effect of anion As height on the electronic structure especially
the Fermi surface of nonmagnetic ThFeAsN.
  
\begin{table}
\begin{center}
\caption{Total energy [$\Delta E$ in meV/f.u., relative to the nonmagnetic state (NM)], density of states 
at the Fermi level [$N(E_F)$ in states/eV/spin/f.u.], total magnetic moment ($m_s^{tot}$ in $\mu_B$/f.u.), 
magnetic moments on transition metal atoms ($m_s^{Fe}, m_s^{Co}$ in $\mu_B$),
the As height ($h_{As}$ in \AA), As ($z_{As}$) and Th ($z_{Th}$) positions 
in ThFeAsN, ThCoAsN and ThNiAsN in the NM,
ferromagnetic (FM), checkerboard antiferromagnetic (c-AF)  and stripe antiferromagnetic (s-AF) states.
Calculated magnetic moments on non-transition metal atoms are less than 0.01 $\mu_B$
and thus are not listed here.
Note that the FM and AF states cannot be not stabilized in ThFeAsN and ThCoAsN, respectively.
Moreover, no FM nor AF states can be stabilized in ThNiAsN.}
 \vspace{2mm}
 \label{mag}
 \begin{tabular}{l c c c c c c} 
 \hline \hline
& \multicolumn{3}{c}{ThFeAsN} & \multicolumn{2}{c}{ThCoAsN} & ThNiAsN \\ \hline
                  & NM & c-AF & s-AF    & NM & FM    & NM \\ \hline
$\Delta E$ & 0  & -14.12 & -82.79 & 0  &-34.28 & 0 \\
  $N(E_F$) & 2.29 & 1.87 & 0.19 & 3.6 & 0.88 (up) & 1.70 \\
           &      &      &     &      & 0.43 (dn) & \\
$m_s^{Fe/Co}$ & 0 & 1.37 & 1.89 & 0 & 0.58 & 0\\
$m_s^{tot}$   & 0 & 0 & & 0 & 1.16  & 0\\
  $z_{As}$ & 0.6388 & 0.6430 & 0.6483 & 0.6404 & 0.6405 & 0.6405 \\
  $z_{Th}$ & 0.1382 & 0.1377 & 0.1374 & 0.1377 & 0.1396 & 0.1396 \\
$h_{As}$ & 1.18  & 1.22 & 1.26 & 1.17 & 1.17 & 1.12   \\ \hline \hline
\end{tabular}
\end{center}
\end{table}

\begin{figure}[h!]
   \centering
   \includegraphics [width=8.5cm]{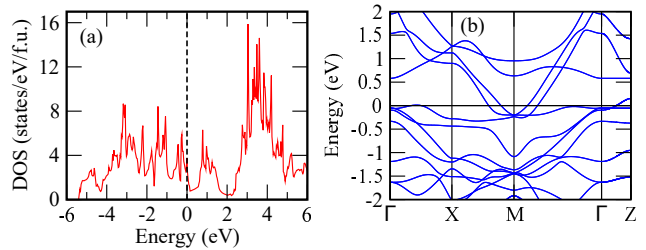}
   \caption{Calculated (a) density of states and (b) band structure of checkerboard antiferromagnetic (c-AF) ThFeAsN.}
   \label{AFMBS}
\end{figure}

\begin{figure}[h!]
   \centering
   \includegraphics [width=8.5cm]{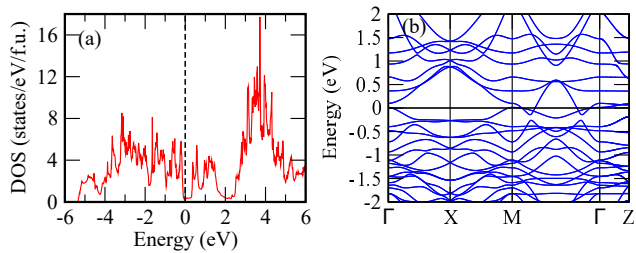}
\caption{Calculated (a) density of states and (b) band structure of stripe antiferromagnetic (s-AF) ThFeAsN.
}
   \label{AFMBS1}
\end{figure}

\begin{figure}[h!]
   \centering
   \includegraphics [width=8.5cm]{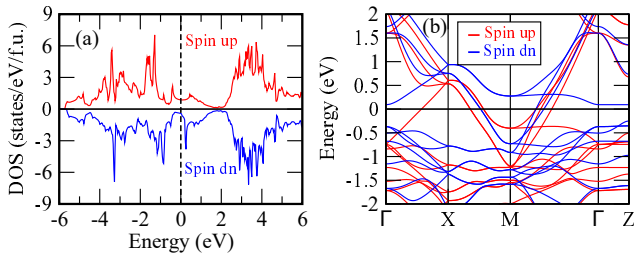}
\caption{Calculated spin-polarised (a) density of states and (b) band structure
of ferromagnetic ThCoAsN.
}
   \label{FMBS}
\end{figure}

\subsection{Magnetism}

Here we consider three possible magnetic structures in ThXAsN (X = Fe, Co, Ni),
i.e., the ferromagnetic (FM) as well as the checkerboard antiferromagnetic (c-AF) 
and stripe antiferromagnetic (s-AF) structures~\cite{2Wang2016}.
The magnetic unit cell of the FM and c-AF structures is the same as the chemical unit cell,
while the magnetic unit cell for the s-AF structure is $\sqrt{2}a\times\sqrt{2}a\times c$
with the Cartesian coordinates rotated 45$^{\circ}$.
Using the lattice constants listed Table I, we perform structural optimization calculations
for these FM and AF states for all the three compounds with the atomic positions fully relaxed, 
as described in Sec. II.
The calculated properties of the magnetic as  well as nonmagnetic ThXAsN (X = Fe, Co, Ni) 
are listed in Table \ref{mag}.
First of all, we find that in ThFeAsN, both c-AF and s-AF states could be stabilized 
with the s-AF state having the lowest total energy (Table \ref{mag}), while the FM state cannot be stabilized. 
This is consistent with previous theoretical calculations~\cite{2Wang2016,Singh2016}.
Nonetheless, as mentioned in Sec. I, no long-range magnetic order has been experimentally 
observed in ThFeAsN~\cite{Wang2016}. This suggests that the GGA calculations may overestimate
the tendency towards the antiferromagnetism in ThFeAsN.
Interestingly, Table \ref{mag} shows that the anion height $h_{As}$ in the s-AF phase (1.26 \AA) 
is significantly larger than that in the NM case (1.18 \AA) and thus 
moves closer to the experimental value of 1.30 \AA~\cite{Mao2017}.
Second, our calculations indicate that ThCoAsN is ferromagnetic and no AF states can be stabilized.
Table \ref{mag} shows that the calculated Fe and Co magnetic moments are significant,
although they are considerably smaller than the Fe and Co magnetic moments
in ferromagnetic Fe and Co metals, respectively. The calculated Th, As and N magnetic moments 
are smaller than 0.01 $\mu_B$ and are thus not listed in Table \ref{mag}.
Finally, we find that no magnetic state can be stabilized in ThNiAsN, i.e., ThNiAsN is nonmagnetic.

The DOS spectra and band structures of the c-AF and s-AF ThFeAsN and FM ThCoAsN 
are plotted in Figs. \ref{AFMBS}, \ref{AFMBS1} and \ref{FMBS}, respectively.
Roughly speaking, the DOS spectra of NM, c-AF and s-AF ThFeAsN are rather similar  
[see Figs. 2(a), \ref{AFMBS}(a) and \ref{AFMBS1}(a)]. In particular, there is a valley near $E_F$.
One significant difference is that the $E_F$ sits on the lower slope of the valley in
NM ThFeAsN, while it is located at the bottom of the valey in the c-AF and s-AF cases. 
This results in a much smaller DOS at $E_F$ in s-AF ThFeAsN (0.19 states/eV/f.u.)
compared with NM ThFeAsN (Table \ref{mag}). Overall, the band structures of c-AF and s-AF  ThFeAsN
are also quite similar, except  that the number of energy bands in  s-AF  ThFeAsN is
twice as many as in c-AF because the number of atoms in the s-AF  ThFeAsN unit cell is doubled.
Nonetheless, the  energy bands in the vicinity of  $E_F$ in NM, c-AF and s-AF ThFeAsN are all quite different.
Figure \ref{FMBS}(b) shows that in FM ThCoAsN, the energy bands are significantly spin-split
with spin-down bands (blue curves) moved upwards and spin-up bands (red curves) shifted-downward. 
Consequently, there is no flat band near $E_F$, and
the DOS spectrum near $E_F$ in FM ThCoAsN [Fig. \ref{FMBS}(a)] differ significantly from that
of NM ThCoAsN [Fig. 2(b)]. In particular, the DOS at the Fermi level in FM ThCoAsN
get much reduced compared with NM ThCoAsN (see Table \ref{mag}).

Since there is theoretical evidence for the AF orders in ThFeAsN especially the SDW-type s-AF order
(Table \ref{mag}), 
we want to discuss the possible effects of the SDW-type AF order on the electron-phonon coupling constant 
and hence superconductivity in ThFeAsN.
It was recently observed that the electron-phonon coupling in some Fe-based SCs increases in the presence of
the AF order~\cite{Deng2009}. It was also shown that the electron-phonon coupling constant could be further increased
when the soft out-of-plane lattice vibrations were taken into account.~\cite{Coh2016}
For a collinear AF system, occupied up and down spin electron orbitals located on a different subset of magnetic atoms in the unit cell.
On the other hand, in the NM state, orbitals of both spin types exist at equal amplitudes on both Fe atoms in the unit cell. When the atoms 
vibrate around their equilibrium positions, the effective potential is changed accordingly, which leads to an enhancement of the electron-phonon 
coupling matrix elements, so that the electron-phonon coupling becomes doubled at the occupied position for this particular
spin orientation \cite{Coh2016}.
According to Ref. ~\cite{arXiv}, one can write the effective electron-phonon coupling on the Fermi surface as
\begin{equation}
\lambda_{eff}=\lambda_{E_F}R_{ph}^2R_{SDW}^2R_{g}^2
\end{equation}
where $R_{ph}$ is the amplification factor due to the soft out-of-plane lattice vibrations
and it turns out to be $~2$ in the AF background \cite{Deng2009}.
$R_{SDW}$ is the amplification factor due to the AF order which is also $~2$ \cite{Coh2016}.
$R_{g}$ is the amplification due to the electrons scattered below the Fermi surface which we
ignore in the present calculation. In the strong coupling case, the renormalized effective electron-phonon coupling
can be written as \cite{arXiv}
$\lambda_{eff}^*=\lambda_{eff}/(\lambda_{eff}+1)$
and the renormalized screened electron-electron repulsion can be written as
$\mu_{eff}^*=\mu_{eff}/(\mu_{eff}+1)$.
After employing all the amplification factors in $\lambda$, we get an effective electron-phonon coupling constant $\lambda_{eff}$
of 0.82 for ThFeAsN. This value of $\lambda_{eff}$ together with the renormalized $\mu$ would result in a $T_c$ of 10.4 K,
being in the same order of magnitude as the experimental $T_c$ value of 30 K. This indicates that an accurate treatment
of the combined effect of the electron-phonon coupling and antiferromagnetism in ThFeAsN would be fruitful,
which, however, is a project for the future.

\begin{figure}
    \centering
    \includegraphics [width=7.0cm]{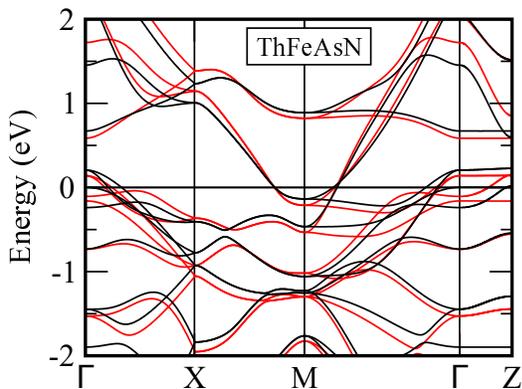}
    \caption{Band structures of nonmagnetic ThFeAsN calculated using experimental
(black lines) and theoretical (red) atomic positions.}
    \label{GBS}
 \end{figure}

 \begin{figure}
    \centering
    \includegraphics [width=8.5cm]{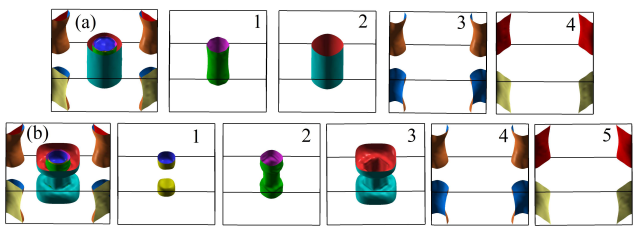}
    \caption{Fermi surfaces of nonmagnetic ThFeAsN calculated using (a) experimental and
(b) theoretical atomic positions.}
    \label{optFS}
 \end{figure}

\subsection{Anion As height and electronic structure}
In this subsection, we examine the effect of the anion height (i.e., the As atomic position 
relative to the Fe-Fe plane $h_{As}$)
on the electronic structure of ThFeAsN. We plot the band structure and Fermi surface of ThFeAsN,
respectively, in Figs. \ref{GBS} and \ref{optFS}, calculated using both the theoretical
and experimental ($z_{As}=0.6522$, $z_{Th}=0.1380$)~\cite{Mao2017} atomic positions
with the fixed experimental lattice constants (see Table I).
Figure \ref{GBS} show that low energy band structures of ThFeAsN calculated using
the theoretical and experimental atomic positions differ significantly.
In particular, the number of bands that cross the Fermi level in the vicinity of the $\Gamma$ point is different.
This results in the different FS shapes and different numbers of the FS pockets for the two cases (Fig. \ref{optFS}).
One pronounced difference is that the FS sheets centered at the $\Gamma$ point calculated
using the experimental atomic positions are much more two-dimensional than that
obtained using the theoretical atomic positions (see Fig. \ref{optFS}).
This would affect the degree of the FS nesting in ThFeAsN.
The FS nesting is believed to play a crucial role in the appearance of
various exotic phases including superconductivity in the Fe based SCs \cite{Sen,Sunagawa2014}.
These pronounced differences in the electronic structure of ThFeAsN are due to
the significant difference in the anion height,
which is caused by the GGA underestimation of the anion height~\cite{2Wang2016, Singh2016}.
Furthermore, our recent calculations~\cite{Sen2020} by using the semilocal DFT-D3 exchange-correlation potential
of Grimme and co-workers~\cite{Grimme2010} to take into account of van der Waals interaction,
gave rise to an anion height that is identical to the GGA value.
Therefore, we speculate that to theoretically obtain the accurate anion height in ThFeAsN,
one should include the effect of magnetic fluctuation, which may play a major role in the physical properties including
superconductivity in ThFeAsN.

\section{CONCLUSIONS}
Summarising, we have carried out a comparative first principles study of mechanical properties,
electronic structure, phonon dispersion relations, electron-phonon coupling and magnetism
in isostructural superconductors ThFeAsN, ThCoAsN and ThNiAsN.
Firstly, our calculated elastic constants indicate that all three compounds are mechanically stable.
Our calculations also reveal that ThFeAsN lies in the boundary of brittle and plastic behaviour, 
being similar to that of the LaFeAsO system. On the other hand,
the Co and Ni systems show rather plastic behaviour. 
The calculated plasma frequencies indicate that ThFeAsN possesses a more 3D-like electronic structure 
than that of the Co and Ni systems.
Secondly, our calculated electronic structures indicate that overall, the three compounds 
have a rather similar electronic structure. Nonetheless, because of their different numbers of
conduction electrons, their Fermi levels are located at rather different energy positions (see Fig. 2).
This result in quite different low-energy band structures in the vicinity of the Fermi level (see Fig. 4).
For example, density of states at the Fermi level $N(E_F)$ for the Co system is twice as large as 
that for the Fe and Ni systems (see Table V).
Topologies of the Fermi surface of the Ni and Co systems are quite similar, 
whereas the FS topology of the Fe system is very similar to that of the LaFeAsO system. 
The FS of ThFeAsN is mostly derived from Fe $d_{z^2}$ and $d_{xz+yz}$ orbitals, whereas for the Co system, 
it is dominated by the $d_{xy}$ orbital.
Thirdly, our calculated phonon dispersions show that these three systems are dynamically stable.
Our electron-phonon coupling calculations indicate that ThCoAsN and ThNiAsN have
nearly the same values of electron-phonon coupling constant $\lambda$ 
while the $\lambda$ value of ThFeAsN is less than half of that for the Co and Ni systems (Table V).
As a result, ThCoAsN has the highest superconducting transition temperature $T_c$ of 6.4 K (Table V).
The $T_c$ for ThNiAsN is 3.4 K, which is quite close to the experimental value of 4.3 K.
This indicates that the superconductivity in ThNiAsN is phonon mediated.
However, ThFeAsN has a nearly zero $T_c$ of 0.05 K, which is far too small compared with
the experimental value of $\sim$30 K (Table V). 

In addition to the systematic calculations for NM ThXAsN (X = Fe, Co, Ni)
mentioned above, we have also studied the effect of anion As height on the low-energy electronic structure of 
NM ThFeAsN and possible magnetic states in all the three compounds.
Interestingly, we find that the c-AF and s-AF structures can
appear in ThFeAsN, although the FM state cannot be stabilized.
Furthermore, the calculated anion height $h_{As}$ in the s-AF state is much closer to
the experimental value than in the NM state (Table VI).
However, no long-range AF order has been observed.
This suggests that the GGA calculations may overestimate the tendency towards the antiferromagnetism in ThFeAsN.
Since there is theoretical evidence for the s-AF order in ThFeAsN (Table \ref{mag}),
we examine the possible effects of the SDW-type AF order on the electron-phonon coupling 
and hence superconductivity in ThFeAsN.
We find an significant enhancement of $\lambda$ when the SDW-type AF background and soft-out-of-plane lattice vibrations 
are taken into consideration. This leads to a $T_c$ of $\sim$10 K.
We thus speculate that the superconductivity in ThFeAsN may be caused by mechanisms other than purely electron-phonon coupling,
such as the SDW-type AF fluctuation mediated-one.
Finally, ThCoAsN is predicted to be ferromagnetic while ThNiAsN is found to be nonmagnetic.
If the ferromagnetism does occur in ThCoAsN, the superconductivity in NM ThCoAsN predicted in Sec. III. will be suppressed 
due to the ferromagnetic pair breaking~\cite{Guo1990}. However, as mentioned above, the GGA calculations
may overestimate the tendency towards the magnetism. Consequently, the predicted ferromagnetic state may not appear
in ThCoAsN and hence the predicted superconductivity may be observed instead. 

\section*{Acknowledgements} 
The authors acknowledge the support from the Ministry of
Science and Technology and National Center for Theoretical Sciences (NCTS) in Taiwan.
The authors are also grateful to the National
Center for High-performance Computing (NCHC) in Taiwan for the computing time.
G.-Y. Guo also thanks the Far Eastern Y. Z. Hsu Science and Technology Memorial 
Foundation in Taiwan for its support.

\end{document}